\documentclass[prd, nofootinbib, preprintnumbers, twocolumn, showpacs]{revtex4}
\usepackage{graphicx,epsf,epsfig}
\begin{document}
\title{Possible non-decoupling effects of heavy Higgs bosons in $e^+ e^- \to W^+ W^-$ within THDM }
\author{Michal Malinsk\'{y} $^{1,2}$}\email{malinsky@sissa.it}
\author{ Ji\v{r}\'{\i} Ho\v{r}ej\v{s}\'{\i}  $^{2}$}\email{horejsi@ipnp.troja.mff.cuni.cz}
\affiliation{
$^{1}$  S.I.S.S.A., Trieste, Italy \\
$^{2}$  IPNP, Faculty of Mathematics and Physics, Charles University, Prague
}
\begin{abstract}
We discuss the origin of the nondecoupling effects of the heavy Higgs bosons within the two Higgs
doublet extension (THDM) of the Standard Model (SM) and illustrate it by means of the one-loop calculation
of
 the differential cross-sections of the process $e^+ e^-
\to W^+ W^-$ in both the decoupling and the non-decoupling regimes.
We argue that there are many regions in the THDM parametric space in which the
THDM and SM predictions differ by several percents and such effects could, at least in principle, be
testable at the future experimental facilities.
\end{abstract}
\pacs{12.15.-y, 12.15.Lk, 12.60.Fr}
\maketitle
\section{Introduction} 
Though being on the market for more than thirty years, the two-Higgs-doublet model (THDM)
\cite{Lee:iz}
still provides one of the most viable extensions of the Standard Model (SM) and, surprisingly 
enough, the activity in this field seems to grow in the last decade -- see e.g. 
\cite{Iltan:2001gf,Krawczyk:2002df,Gunion:2002zf,Kanemura:2002vm,Sanchez}  and
references therein.
 It earns its popularity namely because of its capability to incorporate many sources of  physics
beyond SM \cite{Sher:1988mj,Weinberg:1990me}, be it the CP-violation in the Higgs sector, additional
contributions to the anomalous magnetic moment
of the muon or simply the fact that
the two Higgs doublet structure with 5 massive physical states (and at least one around the weak scale)
mimics nicely many features of the minimal supersymmetric SM (MSSM). 
On top of that, as we shall see, the
Higgs sector of THDM can 
exhibit some particular features which are completely absent in MSSM, namely the relatively large
nondecoupling effects of the heavy Higgs bosons
\cite{Ahn:1988fx,Kanemura:1997wx,Arhrib:1999rg,Ginzburg:1999fb,Ginzburg:2001ph,Malinsky:2002mq,Arhrib:2003ph,Kanemura:2004mg} 
which can arise once the heavy Higgs spectrum is sufficiently nondegenerate. 

The paper is organized as follows: in Section \ref{sect2} we present a general systematic
discussion of the origin of various nondecoupling
effects and the connection of their magnitude with the parameters of the Higgs potential and the
shape of the heavy Higgs spectrum.

In Sections \ref{sect3} and \ref{sect4} we use these results to estimate quantitatively the scale of the
effects in question in the physical process of particular interest, namely $e^{+}e^{-}\to W^{+}W^{-}$.
We thus extend the earlier analysis of S.~Kanemura et al. \cite{Kanemura:1997wx} based on
the equivalence theorem (ET) for longitudinal vector bosons \cite{Cornwall:1974km}, valid in the large $s$
limit.
 Having in hand the results of our previous works \cite{Malinsky:2003bd} and \cite{Malinsky:2003am} in
which we discussed the nondecoupling structures arising in the one-loop triple gauge vertices of THDM
we can go far beyond the ET approximation. 

Section \ref{sect5} is devoted to several quantitative illustrations of the typical behaviour of the
relevant
cross sections both in the decoupling and the nondecoupling regimes, in full agreement with the
estimates.  
 
\section{THDM overview \label{sect2}} 
Adopting the notation of  \cite{Tohyama1}
the most general form of the THDM Higgs potential reads
 $$
V(\Phi_1,\Phi_2) = m_{11}^2\Phi_1^\dagger\Phi_1+m_{22}^2\Phi_2^\dagger\Phi_2-
        (m_{12}^2\Phi_1^\dagger\Phi_2+ {\rm h.c.})+ 
$$
\begin{equation}
\label{potential}
       + \frac{\lambda_1}{2}(\Phi_1^\dagger\Phi_1)^2+
        \frac{\lambda_2}{2}(\Phi_2^\dagger\Phi_2)^2 + 
\end{equation}
$$
        +\lambda_3(\Phi_1^\dagger\Phi_1)(\Phi_2^\dagger\Phi_2)
+\lambda_4(\Phi_1^\dagger\Phi_2)(\Phi_2^\dagger\Phi_1)+ $$       
$$
  +\left\{\frac{\lambda_5}{2}(\Phi_1^\dagger\Phi_2)^2
+[\lambda_6(\Phi_1^\dagger\Phi_1)+\lambda_7(\Phi_2^\dagger\Phi_2)]
        (\Phi_1^\dagger\Phi_2)+ {\rm h.c.}\right\} $$
Here the $\Phi_{1}$ and $\Phi_{2}$ are two $SU(2)$ doublets with the same SM-like hypercharges.
The neutral components of both of them can develop nonzero VEVs around the weak scale and break the 
electroweak symmetry downto the electromagnetic $U(1)$ in the usual manner.

The massive Higgses $h^{0}$, $H^{0}$, $A^{0}$ and $H^{\pm}$ as well as the Goldstone bosons $G^{0}$ and
$G^{\pm}$
are then 'encoded' within these doublets as
\begin{eqnarray}
\Phi_1 &\!\! =\!\! &
\frac{1}{\sqrt{2}}\left[\!
\begin{array}{c}
    \sqrt{2}G^+\cos
\beta - \sqrt{2}H^+\sin
\beta
\\
H^0\cos\alpha-h^0\sin\alpha+v_1+iG^0\cos
\beta-iA^0\sin
\beta
    \end{array}\right] \nonumber \\
    \Phi_2 &\!\! =\!\! &
\frac{1}{\sqrt{2}}\left[\!
\begin{array}{c}
    \sqrt{2}G^+\sin
\beta + \sqrt{2}H^+\cos
\beta
\\
H^0\sin\alpha+h^0\cos\alpha+v_2+iG^0\sin
\beta+iA^0\cos
\beta
    \end{array}\right]
\nonumber
\end{eqnarray}
where as usual $\tan
\beta=v_2/v_1$. For the sake of simplicity we take both $v_1$ and $v_{2}$ real. 
Next, the neutral scalar mixing angle $\alpha$ is given by 
\begin{equation}
\label{cosbetaalpha}
\cos^2(\alpha-
\beta)=\frac{m_{L}^2-m_{h^{0}}^2}{m_{H^0}^2-m_{h^0}^2} 
\end{equation}
where we have denoted
$$
m_L^2\equiv \left[\lambda_1\cos^4
\beta+\lambda_2\sin^4
\beta+\frac{1}{2}\left(\lambda+2D\sin^2\beta\right)\right]v^2 $$
with $\lambda\equiv \lambda_{3}+\lambda_{4}+\lambda_{5}$ and $D  \equiv  \lambda_7^R \tan
\beta + \lambda_6^R \cot
\beta $ (the superscript $R$ denotes the real part).
\subsection{The physical Higgs spectrum \label{sect2-1}} 
Let us inspect the general formulae for the physical Higgs masses descending from the potential
(\ref{potential}) :
\begin{eqnarray}
\label{masses} m_{h^0}^2 & = & \frac{1}{2}(1-\kappa)M^2+\frac{v^2}{\cos 2\alpha }\times \nonumber \\
& &\times \left[B_2\sin^2
\beta - A_1\cos^2
\beta + \frac{1}{4}C (1+\cos 2\alpha \cos 2\beta)\right]\nonumber \\
m_{H^0}^2 & = & \frac{1}{2}(1+\kappa)M^2+\frac{v^2}{\cos 2\alpha }\times \nonumber  \\
& & \times \left[A_2\cos^2
\beta -B_1\sin^2
\beta - \frac{1}{4}C (1-\cos 2\alpha \cos 2\beta)\right] \nonumber \\
m_{A^0}^2 & = & M^2 -\frac{1}{2}\left(2\lambda_5^R+\lambda_6^R\cot\beta + \lambda_7^R\tan
\beta\right) v^2  \\
m_{H^\pm}^2 & = & M^2 -\frac{1}{2}\left(\lambda_4+\lambda_5^R+\lambda_6^R\cot\beta +
\lambda_7^R\tan\beta\right) v^2 \nonumber 
\end{eqnarray}
where 
\begin{eqnarray}
M^2 & \equiv & \frac{{m^2_{12}}^R}{\sin
\beta \cos
\beta}\qquad\qquad  
\kappa \equiv -\frac{\cos 2
\beta}{\cos 2\alpha}\nonumber \\
A_1 & \equiv & \lambda_1 \sin^2\alpha - \lambda_7^R \tan
\beta \cos^2\alpha  \nonumber \\
A_2 & \equiv & \lambda_1 \cos^2\alpha - \lambda_7^R \tan
\beta \sin^2\alpha \nonumber \\
B_1 & \equiv & \lambda_2 \sin^2\alpha - \lambda_6^R \cot
\beta \cos^2\alpha \nonumber \\
B_2 & \equiv & \lambda_2 \cos^2\alpha - \lambda_6^R \cot
\beta \sin^2\alpha\nonumber \\
C  & \equiv & \lambda_7^R \tan
\beta - \lambda_6^R \cot
\beta  \nonumber
\end{eqnarray} 
Notice that the two mass parameters $m_{11}^2$ and  $m_{22}^2$ were as usual fixed by the necessary conditions
for the VEVs of $\Phi_1$ and $\Phi_2$ to minimize the potential. 
For $\lambda_{6}=\lambda_{7}=m_{12}=0$ 
one recovers the formulae given previously in
the literature, see e.g. \cite{Tohyama1}, \cite{Tohyama2} and references therein.

Let us call by {\it heavy Higgs mass limit}
the situation, in which the masses of all the THDM Higgs bosons but $h^0$ are much larger than the
weak scale. 

One can see that there are in general two basic quantities responsible for the shape of
the Higgs spectrum (\ref{masses}): the
gauge singlet mass parameter $M$ (alias $m_{12}$)  and the VEV magnitude $v$. Since $M$ is not protected
by the
gauge  symmetry it could be naturally much larger than $v$ and in such case the heavy Higgs mass limit
is achieved entirely by enlarging $M$.
On the other hand, there are unitarity
bounds on the masses of the 'heavy' members of the Higgs spectrum preventing them to be extremely heavy
\cite{Huffel}.
Moreover, the $v$ is often accompained by (in principle) numerically large factors
$\propto \lambda_{7}\tan \beta$ (or $\lambda_{6}\cot \beta)$ which enhances some of the '$\lambda v^{2}$
terms' obviating to large
extent the necessity of having a dominant $M$ to achieve the heavy Higgs mass limit.

This is
in sharp contrast with the situation in the MSSM where only one free parameter $\mu$ is left to play
with, because the quartic couplings in the Higgs potential are fixed by supersymmetry.

\subsection{\label{nondecoupling}\label{sect2-2}Nondecoupling regime \& spectrum distorsions} 
Therefore, it
is convenient to distinguish between two different modes in which the 'heavy part' of the Higgs spectrum
acquires the masses: 
\begin{itemize}
\item {if it is due to the dominance of the singlet mass terms ($M$-components) in the relations
(\ref{masses}) let us call it {\it the decoupling regime}. Perhaps it is worth noting that although
the only explicit mass present in (\ref{masses}) is $m_{12}$ one should not forget about 
$m_{11}$ and $m_{22}$ that are  'hidden' in particular combinations of the other parameters in the
game which can to some extent mimic their role unless $\lambda_{6,7}=0$ (see also comments in Section
\ref{sect5-2}) }.
\item {the contributions coming from the $M$ components are comparable with the other '$\lambda v^{2}$' parts in
(\ref{masses});  such situation is called {\it the nondecoupling regime}.}
\end{itemize} 
As the terminology suggests, in the decoupling regime the heavy Higgs bosons exhibit a decoupling
behaviour in
accordance with the famous Appelquist-Carazzone theorem \cite{Appelquist:tg}. In this case one can easily
show
that the requirement of coincidence of the THDM $h^0$ with the SM Higgs boson $\eta$ (with masses not
far from
$m_W$) and the relation (\ref{cosbetaalpha}) lead to $\kappa\sim 1$ and therefore {\it the heavy
Higgs spectrum is quasidegenerate $m_{H^0}\sim $ $m_{A^0}\sim $ $m_{H^\pm}\sim M$}.

On the other hand, in the nondecoupling regime {\it the heavy Higgs spectrum
should be distorted} and one
can in principle expect substantial effects in measurable quantities which should grow with the weights
of the $\lambda v^{2}$ terms in (\ref{masses}), i.e.
with the magnitude of such a distortion. 
This was used as a nontrivial consistency check of the numerical results we present
below.

From this point of view the behaviour of the heavy Higgs bosons in the MSSM is very
simple in comparison with THDM; in fact, in MSSM there is no such nondecoupling regime and all the heavy
Higgses therein should therefore tend to decouple from the weak scale physics. This was
confirmed explicitly in \cite{Dobado:2000pw}.
 
This can be used e.g. as a simple heuristic explanation of what happens in
\cite{Tohyama1} and
\cite{Tohyama2}, where the considered
non-decoupling effects of the heavy part of the Higgs spectrum tend to minimize provided a partial
degeneration in the heavy Higgs sector is achieved (
$m_{H^{\pm}}\to m_{A^{0}}$). 

In the remaining part of the paper we demonstrate these principles in the particular case of
physical cross-sections of the process $e^- e^+ \to W^- W^+$ computed within the THDM framework at
one-loop order in comparison with the SM predictions. We generalize the earlier
work \cite{Kanemura:1997wx} beyond the ET approximation used therein.
Among other things, our approach allows one to consider general configurations of polarizations of the
final state vector bosons.
%

\section{\label{sect3}${\rm d}\sigma(e^- e^+ \to W^- W^+)$ in THDM versus SM} Since the one-loop form of
the differential cross-section within the SM  is very well known \cite{Bohm:1987ck}, we can use the
similarity of THDM to simplify our life by dealing with the pieces of information which are specific
for THDM,
namely the contributions to the one-loop amplitude and the bremsstrahlung terms that are  different  for
THDM and SM. 

Therefore, it is reasonable to work with a quantity that measures just the deviation of
the THDM and SM cross sections under consideration \cite{Kanemura:1997wx}; let us define it as
\begin{equation} \label{delta}
\delta \equiv 
\frac{{\rm d}\sigma^{{THDM}}(e^+e^- \to W^+W^-) } {{\rm d}\sigma^{{SM}}(e^+e^- \to W^+W^-) } - 1 
\end{equation}
Both differential cross section in the last expression can be written as     
$$
{\rm d}\sigma={\rm d}\sigma_{A}+{\rm d}\sigma_{B} $$
where the pieces ${\rm d}\sigma_{A}$ come from the 'amplitude-squared' terms 
\begin{equation}
{\rm d}\sigma_{A}=k_1\left|{\cal M}\right|^2 {\rm d}Lips
\end{equation}
 while the ${\rm d}\sigma_{B}$ terms represent the bremsstrahlung effects 
\begin{equation}
{\rm d}\sigma_{B}=k_2\int {\rm d}k_\gamma \left|{\cal B}(k_\gamma,\ldots)\right|^2 {\rm d}Lips
\end{equation}
Expanding now the THDM amplitude around the corresponding SM form
\begin{eqnarray}
{\cal M}^{THDM}_{tree} & = & {\cal M}^{SM}_{tree}+\Delta{\cal M}_{tree} \nonumber \\
{\cal M}^{THDM}_{loop} & = & {\cal M}^{SM}_{loop}+\Delta{\cal M}_{loop} \nonumber
\end{eqnarray}
($\Delta{\cal M}_{tree}$ and $\Delta{\cal M}_{loop}$ are just the differences of the tree- and
one-loop amplitudes respectively) one can recast the $\delta$ as
\begin{equation}
\label{firststage}
\delta = 2{\rm Re}\left[\frac{\Delta{\cal M}_{{tree}}+\Delta{\cal M}_{{loop}}} {{\cal
M}_{{tree}}^{{SM}}}\right]+
\end{equation}
$$
+\frac{k_2}{k_1}
\int {\rm d}k_\gamma
\frac{
\left|{\cal B}^{THDM}\right|^2-\left|{\cal B}^{SM}\right|^2}{\left|{\cal M}_{tree}^{SM}\right|^2} +
\ldots 
$$
Here the $k_i$ are ${\cal O}(1)$ geometrical factors and the integration over  $k_\gamma$ covers the
IR-singular piece of the phase space spanned by the photon momenta.  
\subsection{\label{sect3-1}Bremsstrahlung terms} Let us first explore the bremsstrahlung terms in
(\ref{firststage}).
Using the identity $|A|^2-|B|^2=\frac{1}{2}(A-B)(A+B)^*+h.c.$ we see that the only terms that survive
(i.e. those that are not common to both THDM and SM) come from the graphs ($H$ denotes the generic neutral
Higgses in the game): $$
\parbox{25mm}{\epsffile{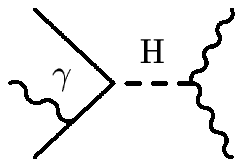}}
\qquad{\rm and}\qquad
\parbox{25mm}{\epsffile{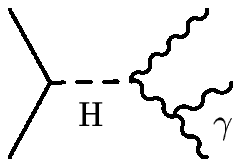}}
$$
These diagrams are necessary to regulate the singular infrared behaviour of ${\rm d}\sigma_{A}$ caused by
the presence of $$
\parbox{25mm}{\epsffile{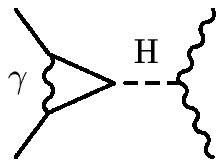}}
\qquad{\rm and}\qquad
\parbox{25mm}{\epsffile{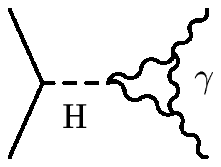}} $$
All the above graphs are IR singular but the divergences cancel in the physical cross-section formulae
and once we specify the IR regulator scale they provide just logarithmic corrections to ${\rm
d}\sigma_{A}$. As we shall see, the presence of the electron Yukawa coupling makes these terms
suppressed
by factors of the order of $m_e/m_W \sim 10^{-5}$ with respect to the leading contributions to $\delta$
coming from
the IR-regular terms in ${\rm d}\sigma_{A}$. Thus, in the leading approximation we can omit the
bremsstrahlung effects entirely.

\subsection{\label{sect3-2}The leading term} This means that we can forget about the second
term of the expansion
(\ref{firststage}) and also keep only the IR-regular subset of graphs corresponding to ${\cal
M}^{THDM}_{loop}$, which we shall denote by $\overline{{\cal M}}^{THDM}_{loop}$. 
Next, it is worth noting that the $\Delta{\cal M}_{tree}$ in (\ref{firststage}) also suffers of the
'omnipresent' electron Yukawa coupling which again makes it much smaller in comparison with the leading
loop contributions to ${{\cal M}}^{THDM}_{loop}$ coming (as we shall see) from the renormalization of the
triple gauge vertices (TGV).
To conclude, the leading contribution to $\delta$ can be written in the form
\begin{equation}
\label{central_relation}
\delta =
 2{\rm Re}\frac{\Delta\overline{{\cal M}}_{loop}}{{\cal M}_{tree}^{SM}}+\ldots
\end{equation}

\section{\label{sect4}Computation of $\Delta\overline{{\cal M}}_{loop}$}
Thus, all we need is only the sum of all the IR-safe one-loop diagrams which are  not shared by the THDM
and SM, i.e. those loop graphs which contain at least one Higgs propagator.

\subsection{\label{sect4-1}Renormalization scheme and gauge choice} 
We have decided to use the on-shell renormalization scheme \cite{Aoki:ed}
in which the renormalized masses parametrizing the one-loop quantities
coincide with the physical masses of the fields in the game; this simplifies greatly the interpretation of
various limits under consideration (in particular, we use the set of counterterms defined in \cite{Pokorski:ed}).
At first glance, it might seem that 
this also obviates
 the necessity to deal with the renormalization of the external legs of Feynman 
graphs. In fact, this is not true because in the on-shell scheme nontrivial finite parts of
various counterterms reappear and these bring back the self-energy diagrams. We will work in 
the Feynman ($\xi=1$) gauge to have the IVB propagators as simple as possible. This of course
requires to take into account also the unphysical Goldstone bosons.

Note that there is no need to take care about the ghost fields (which do not 
decouple from the Higgs sector unless $\xi=0$) because 
all the relevant topologies contributing to 
${\Delta\overline{{\cal M}}_{{loop}}}$ containing the ghost loop involve also the Yukawa couplings
(1-loop irreducible graphs) or do not contribute 
substantially (oblique corrections). 

\subsection{\label{sect4-2}Feynman diagrams contributing to ${\Delta\overline{{\cal M}}_{{loop}}}$ } 
Let us first classify the topologies of Feynman diagrams contributing 
to ${\Delta\overline{{\cal M}}_{{loop}}}$ that do not subtract trivially in (\ref{delta}),
i.e. those that are not common to the two models. 
In what follows, the symbol $\parbox{5mm}{\epsffile{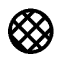}}$
denotes a loop involving at least 
one Higgs propagator while $
\parbox{5mm}{\epsffile{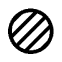}}$ is used for loops built entirely from the other SM fields. 
In both cases also the corresponding counterterms must be taken into account.

Concerning the tree-level origin of the the relevant one-loop graphs one can identify several subgroups
of them, namely:
\paragraph{Neutrino in the $t$-channel:} $$
a)
\parbox{25mm}{\epsffile{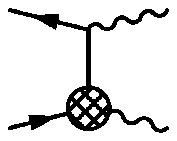}} 
b)
\parbox{25mm}{\epsffile{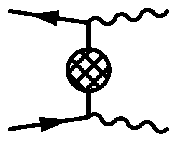}} $$
\paragraph{$Z$ and $\gamma$ in the $s$-channel:} $$
c)
\parbox{25mm}{\epsffile{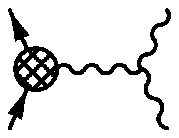}} d)
\parbox{25mm}{\epsffile{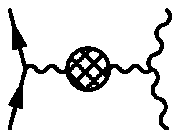}} e)
\parbox{25mm}{\epsffile{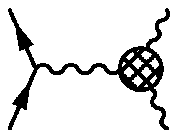}} $$
\paragraph{Higgs in the $s$-channel:}
$$
f)
\parbox{25mm}{\epsffile{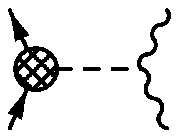}} g)
\parbox{25mm}{\epsffile{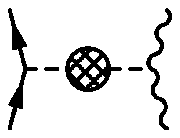}} h)
\parbox{25mm}{\epsffile{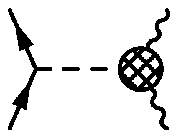}} $$
$$
i)
\parbox{25mm}{\epsffile{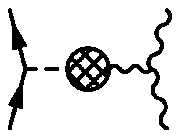}} j)
\parbox{25mm}{\epsffile{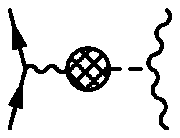}} $$
$$
k)
\parbox{25mm}{\epsffile{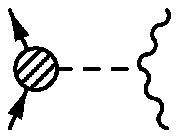}} l)
\parbox{25mm}{\epsffile{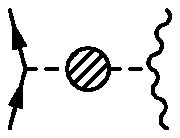}} m)
\parbox{25mm}{\epsffile{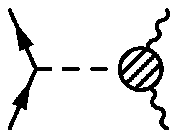}} $$
$$
n)
\parbox{25mm}{\epsffile{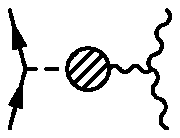}} o)
\parbox{25mm}{\epsffile{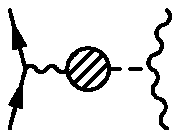}} $$
\paragraph{Box diagrams:} $$
p)
\parbox{25mm}{\epsffile{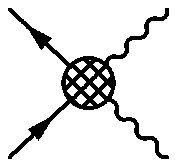}} $$
The total number of the graphs with the topologies displayed here is enormous. However, not all of them
are relevant in the leading approximation. 

\subsection{\label{sect4-3}Relevant topologies} As before, the presence of the electron Yukawa couplings
allows one to
neglect all graphs of the types $a)$, $b)$, $c)$, $f)$, $g)$, $h)$, $i)$, $l)$, $m)$ and $n)$ in
comparison with the types $d)$, $e)$, $k)$, $o)$ and $p)$ where there is no such factor. Next, the blob 
in the type $k)$ provides just a correction to the Yukawa vertex which is fixed by the renormalization
conditions to be comparable with the Yukawa coupling itself. The diagrams of type $o)$ are again
$m_{e}/m_{W}$ suppressed; the electron mass here arises from the Dirac equation.
We are therefore left with the
IVB 'vacuum polarization' graphs in $d)$ (the oblique corrections to the IVB propagators), the corrections
to the triple gauge vertices $e)$ and the UV
convergent box diagrams $p)$. However, among them only those involving the Higgs bosons coupled to the IVB
lines need to be taken into account, otherwise the
small Yukawa coupling reappears. Next, since the boxes are UV-finite they must decouple trivially in
the heavy
Higgs mass limit described above. All that remains at the leading order are therefore the graphs of the
type 
$$
i)\quad
\parbox{25mm}{\epsffile{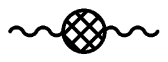}} 
ii)\quad
\parbox{25mm}{\epsffile{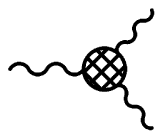}} $$
that we shall treat separately.
 
\subsection{\label{sect4-4}Decoupling behaviour of the oblique corrections} The renormalized inverse
propagator of a massive vector boson is given by
\begin{eqnarray}
i\Gamma^{(1)}_{\mu\nu}(k) & = & i\Gamma^{(0)}_{\mu\nu}(k)+i\Pi_{\mu\nu}(k)+i(Z-1)k^2 P^T_{\mu\nu} -
\nonumber \\
&& -i\delta m^2 g_{\mu\nu}+ik^2P^L_{\mu\nu}\delta\alpha^{-1}
\end{eqnarray}
where $P^T_{\mu\nu}\equiv g_{\mu\nu}-k_\mu k_\nu{k^{-2}}$ and 
$P^L_{\mu\nu}\equiv k_\mu k_\nu k^{-2} $ 
are the transverse and longitudinal projectors, $$i\Pi_{\mu\nu}(k)\equiv ik^2\Pi^T(k^2)P^T_{\mu\nu} +
ik^2\Pi^L(k^2)P^L_{\mu\nu}$$
is the sum of all  
relevant one-loop graphs, $(Z-1)$ and $\delta m^2$ are the 
wave-function and mass counterterms and $\Gamma^{(0)}_{\mu\nu}= 
(k^2-m^2)P^T_{\mu\nu}+\alpha^{-1}\left(k^2-\alpha 
m^2\right)P^L_{\mu\nu}$ 
is the tree-level massive gauge boson propagator. 
In the on-shell scheme the counterterms are fixed by \cite{Aoki:ed}
\begin{equation}
\Gamma^T(k^2=m^2)=0 \quad  
\frac{d}{dk^2}\Big|_{k^2=m^2}{\Gamma^T}(k^2)=1 
\end{equation}
which yields
\begin{eqnarray}
Z-1 & = & -\left[\Pi^T(m^2)+m^2{\Pi^T}'(m^2)\right] \nonumber \\
\delta m^2 & = & -m^4 {\Pi^T}'(m^2) \nonumber
\end{eqnarray}
We do not need to deal with the longitudinal part of the 
renormalized IVB propagators because they typically produce the suppressing $m_e/m_{W}$ factors. 
We can also omit all 
tadpole diagrams: their contributions to $\Pi^T$ are proportional to 
$k^{-2}$ and therefore cancel (as they should) in $Z-1$ and $[k^2\Pi^T(k^2)-\delta m^2]$ in 
$\Gamma^T$.

The remaining graphs are those appearing in the massive scalar 
electrodynamics. It is already easy to show that such contributions to $\delta$
fall rapidly in the heavy Higgs mass limit regardless of the way the limit is achieved. 

\subsection{\label{sect4-5}Triple gauge vertex corrections} 
Concerning the differences of the heavy Higgs corrections to the triple gauge vertices in THDM and in
SM the major part of the work
has already been done before \cite{Malinsky:2003bd}, so let us mention just the important
points.
The differences of the relevant vertex functions can be written in the form 
\begin{eqnarray}
  \label{TGVdifference1}
  \Delta\Gamma^{VWW}_{\sigma\mu\nu} &  = &\sum_{i=1}^3\left(\Delta\delta Z_{TGV}+
\Delta\Pi_i^{VWW}\right)C^i_{\sigma\mu\nu}+ \nonumber
\\
  & & +\sum_{i=4}^7 \Delta\Pi_i^{VWW}C^i_{\sigma\mu\nu}+sym. \nonumber 
\end{eqnarray}
where $$
  C^{1}_{\sigma\mu\nu}   \equiv  {q_1}_\sigma g_{\mu\nu} \qquad
  C^{2}_{\sigma\mu\nu}   \equiv  2q_{2\mu} g_{\sigma\nu} \qquad
  C^{3}_{\sigma\mu\nu}   \equiv  q_{1\mu}  g_{\sigma\nu} $$
\begin{eqnarray}
\label{basicstructures2}
  C^{4}_{\sigma\mu\nu}   \equiv  \frac{1}{m^2_{W}}{q_1}_\sigma q_{1\mu} q_{1\nu} \qquad
  C^{5}_{\sigma\mu\nu}   \equiv  \frac{1}{m^2_{W}}{q_1}_\sigma q_{1\mu} q_{2\nu} \\
  C^{6}_{\sigma\mu\nu}   \equiv  \frac{1}{m^2_{W}}q_{1\sigma} q_{2\mu} q_{1\nu} \qquad
  C^{7}_{\sigma\mu\nu}   \equiv  \frac{1}{m^2_{W}}q_{2\sigma} q_{1\mu} q_{1\nu} \nonumber
\end{eqnarray}
are basic kinematical structures composed of the outgoing momenta $q_{1,2}$ of the $W^{\pm}$ bosons in the
final state. Next,   
\begin{eqnarray}
\label{Zdifferences}
 \Delta \delta Z_{TGV} & \equiv &(\delta Z_{TGV})_{THDM}-(\delta Z_{TGV})_{SM} 
\end{eqnarray}
are the differences of the corresponding finite parts of counterterms $Z_{TGV}$ (computed by means of the
$W$ self-energy diagrams) and 
\begin{eqnarray}
\label{PIdifferences}
 \Delta \Pi_i^{VWW} & \equiv & (\Pi_i^{VWW})_{THDM}-( \Pi_i^{VWW})_{SM}
\end{eqnarray}
are the differences of the formfactors $\Pi_i^{VWW}$ (descending from the triangle diagrams
contributing at the one-loop level to the triple gauge vertices); for more details see
\cite{Malinsky:2003bd}. 

\subsection{\label{sect4-6}$\delta$ revised} 
Armed by the information given in the last three subsections we are ready
to write down the explicit formulae for the leading part of (\ref{central_relation}): $$
{\Delta\overline{{\cal M}}_{{loop}}}\doteq 
\sum_{{V}=\gamma,{Z}} g_{ee{V}}g_{{VWW}}\bar{v}(p_1)\gamma_\lambda 
u(p_2)\frac{-ig^{\lambda\sigma}}{s-m^2_{{V}}}\times $$
\begin{equation}
\label{central_relation2}
\times\Delta\Gamma^{{VWW}}_{\sigma\mu\nu}(q_1,q_2)\,
\varepsilon^{*\mu}(q_1)\varepsilon^{*\nu}(q_2)
\end{equation}
Inspecting (\ref{basicstructures2}) and using the identities
$q_1.\varepsilon(q_1)=q_2.\varepsilon(q_2)=0$, we can write (denoting by $A$ the set
$(q_1^2,q_2^2,q_1.q_2)$ and correspondingly $B \equiv (q_2^2,q_1^2,q_2.q_1)$ , which is nothing but the
'$sym.$' operation applied on $A$, see \cite{Malinsky:2003bd})
\begin{eqnarray}
& & \Delta\Gamma^{{VWW}}_{\sigma\mu\nu}(q_1,q_2)\,
\varepsilon^{*\mu}(q_1)\varepsilon^{*\nu}(q_2)=\varepsilon^{*\mu}(q_1)\varepsilon^{*\nu}(q_2)\times
\nonumber \\
&  & 
\times 
\Bigl[ (\Delta\Pi_1^{VWW}+\Delta\delta Z_{TGV})(A){q_1}_\sigma g_{\mu\nu}-\nonumber \\
& & -(\Delta\Pi_1^{VWW}+\Delta\delta Z_{TGV})(B){q_2}_\sigma g_{\mu\nu}+ 
\nonumber \\
& & + (\Delta\Pi_2^{VWW}+\Delta\delta Z_{TGV})(A)2{q_2}_\mu g_{\sigma\nu}- \nonumber \\
& & -(\Delta\Pi_2^{VWW}+\Delta\delta Z_{TGV})(B)2{q_1}_\nu g_{\sigma\mu}+ \nonumber \\
& & + \left. 
\Delta\Pi_6^{VWW}(A)\frac{{q_1}_\sigma{q_2}_\mu{q_1}_\nu}{m_W^2}-\Delta\Pi_6^{VWW}(B)\frac{{q_2}_%
\sigma{q_1}_\nu{q_2}_\mu}{m_W^2}\right].
\nonumber
\end{eqnarray}
If, for simplicity, we take the final state $W$ bosons on the mass shell, i.e. $q_1^2=q_2^2=m_W^2$ we
get $A=B$ and (\ref{central_relation2}) can be recast in the form
\begin{eqnarray}
& & {\Delta\overline{{\cal M}}_{{loop}}} \doteq 
\sum_{{V}=\gamma,{Z}} g_{ee{V}}g_{{VWW}}\bar{v}(p_1)\gamma_\lambda 
u(p_2)\frac{-ig^{\lambda\sigma}}{s-m^2_{{V}}}\times \nonumber \\
& & 
\Bigl[ (\Delta\Pi_1^{VWW}+\Delta\delta Z_{TGV})(m_W^2,s)({q_1}-{q_2})_{\sigma} g_{\mu\nu}+ \nonumber \\
& & + 2(\Delta\Pi_2^{VWW}+\Delta\delta Z_{TGV})(m_W^2,s)({q_2}_\mu g_{\sigma\nu}-{q_1}_\nu
g_{\sigma\mu})+ \nonumber 
\\
& & + 
\Delta\Pi_6^{VWW}(m_W^2,s)\frac{{q_2}_\mu{q_1}_\nu}{m_W^2}({q_1}-{q_2})_{\sigma}\Bigr]
\varepsilon^{*\mu}(q_1)\varepsilon^{*\nu}(q_2)
\nonumber
\end{eqnarray}
Next, the momentum conservation $q_1+q_2+p_1+p_2=0$ and the Dirac equations
$\slash\!\!\!\!p_i=m_e$ allow for further simplifications. Plugging in the coupling constants given in
\cite{Malinsky:2003bd} one finally arrives at
\begin{eqnarray}
& & {\Delta\overline{{\cal M}}_{{loop}}^{LR}} \doteq 
-i e^2 2\Bigl\{\frac{1}{s}\Bigl[ {\cal M}_2^+\Bigl(\Delta\Pi_1^{\gamma WW}+\Delta\delta Z_{TGV}\Bigr)-
\nonumber \\
& & - {\cal M}_3^+\Bigl(\Delta\Pi_2^{\gamma WW}+\Delta\delta Z_{TGV}\Bigr) +\frac{1}{m_W^2}{\cal
M}_5^+\Delta\Pi_6^{\gamma WW}\Bigr]-\nonumber \\
& & -\frac{1}{s-m_W^2}\Bigl[ {\cal M}_2^+\Bigl(\Delta\Pi_1^{ZWW}+\Delta\delta Z_{TGV}\Bigr)-
\label{first} \\
& & -{\cal M}_3^+\Bigl(\Delta\Pi_2^{ZWW}+\Delta\delta Z_{TGV}\Bigr)+\frac{1}{m_W^2}{\cal
M}_5^+\Delta\Pi_6^{ZWW}
\Bigr]\Bigr\}
\nonumber
\end{eqnarray}
and
\begin{eqnarray}
& & {\Delta\overline{{\cal M}}_{{loop}}^{RL}} \doteq 
-i e^2 2\Bigl\{\frac{1}{s}\Bigl[ {\cal M}_2^-\Bigl(\Delta\Pi_1^{\gamma WW}+\Delta\delta Z_{TGV}\Bigr)-
\nonumber \\
& & - {\cal M}_3^-\Bigl(\Delta\Pi_2^{\gamma WW}+\Delta\delta Z_{TGV}\Bigr) +\frac{1}{m_W^2}{\cal
M}_5^-\Delta\Pi_6^{\gamma WW}\Bigr]-\nonumber \\
& & -\frac{c_\theta}{s_\theta}g_e^-\frac{1}{s-m_W^2}\Bigl[ {\cal
M}_2^-\Bigl(\Delta\Pi_1^{ZWW}+\Delta\delta Z_{TGV}\Bigr)-  \\
& & -{\cal M}_3^-\Bigl(\Delta\Pi_2^{ZWW}+\Delta\delta Z_{TGV}\Bigr)+\frac{1}{m_W^2}{\cal
M}_5^-\Delta\Pi_6^{ZWW}
\Bigr]\Bigr\}
\nonumber
\end{eqnarray}
where we have employed the notation of \cite{Bohm:1987ck}, namely
\begin{eqnarray}
{\cal M}_2^+ & \equiv &
\bar{v}_L(p_1)\,\,\slash\!\!\!\!q_1 u_R(p_2) \varepsilon^*(q_1).\varepsilon^*(q_2)
 \nonumber \\
{\cal M}_3^+ & \equiv & \bar{v}_L(p_1)[\,\,\slash\!\!\! \varepsilon^*(q_1) q_1.
\varepsilon^*(q_2)-\,\,\slash\!\!\! \varepsilon^*(q_2) q_2. \varepsilon^*(q_1)] u_R(p_2)
\nonumber \\
{\cal M}_5^+ & \equiv &\bar{v}_L(p_1)\,\,\slash\!\!\!\!q_1 u_R(p_2) [q_2.\varepsilon^*(q_1)]
[q_1.\varepsilon^*(q_2)]
 \label{invariants} \\
{\cal M}_2^- & \equiv & 
\bar{v}_R(p_1)\,\,\slash\!\!\!\!q_1 u_L(p_2) \varepsilon^*(q_1).\varepsilon^*(q_2)
\nonumber \\
{\cal M}_3^- & \equiv & \bar{v}_R(p_1)[\,\,\slash\!\!\! \varepsilon^*(q_1) q_1.
\varepsilon^*(q_2)-\,\,\slash\!\!\! \varepsilon^*(q_2) q_2. \varepsilon^*(q_1)] u_L(p_2)
\nonumber \\
{\cal M}_5^- & \equiv &
\bar{v}_R(p_1)\,\,\slash\!\!\!\!q_1 u_L(p_2) [q_2.\varepsilon^*(q_1)] [q_1.\varepsilon^*(q_2)]
 \nonumber
\end{eqnarray}
and 
$$
g_e^-\equiv \frac{2s^2_\theta-1}{2s_\theta c_\theta} $$
with $\theta$ being the weak mixing angle. The LR and RL superscripts above denote the helicity
configurations of the initial $e^+e^-$ state. 

The last missing piece is the tree-level amplitude 
${\cal M}^{SM}_{tree}$ that can be recast in terms of these quantities as :
\begin{eqnarray}
& & {{\cal M}_{tree}^{SM,LR}} = 
-i e^2 2\Bigl(\frac{1}{s}-\frac{1}{s-m_W^2}\Bigr)[{\cal M}_2^+-{\cal M}_3^+]
\nonumber \\ \label{SMtreeamplitude}
\end{eqnarray}
and 
\begin{eqnarray}
{{\cal M}_{tree}^{SM,RL}}&  = & -i e^2 
\Bigl\{ 2\Bigl(\frac{1}{s}-\frac{c_\theta}{s_\theta}g_e^-\frac{1}{s-m_W^2}\Bigr)[{\cal M}_2^--{\cal
M}_3^-]- \nonumber \\
& & -\frac{1}{2s^2_\theta}\frac{1}{t}{\cal M}_1^-\Bigr\}
\label{last}
\end{eqnarray}
(cf.~\cite{Bohm:1987ck}).
Substituting now (\ref{first})-(\ref{last}) into (\ref{central_relation}) one can calculate $\delta$ for
both the leptonic helicity configurations. The polarizations of the final states are encoded in the
invariants (\ref{invariants}).

\section{\label{sect5}Results and further comments}
We shall present our results obtained in two 'maximally' different situations: first, in the non-decoupling
regime with very hierarchical heavy Higgs spectrum and then in the decoupling regime with almost
degenerate heavy Higgs masses. 

Our input parameters in both cases are the Higgs masses in the 
game ($m_{\eta}$, $m_{h^0}$, $m_{H^0}$, $m_{A^0}$ and $m_{H^\pm}$).
Each such set fixes four THDM parameters out of 
$m_{12}$, $\lambda_{1..7}$, $\beta$ and $\alpha$; 
all the remaining ones are left to be chosen.
For the sake of simplicity we shall take 
\begin{equation}
\label{lambdas}
\lambda_1 = \lambda_2 \equiv \lambda_{12}
\end{equation}
\subsection{\label{sect5-1}Nondecoupling regime } 
\begin{figure}[h]
\caption{\label{obr1}$\delta$ as a function of $m_H^0=20\Lambda$, $m_{A^0}=10\Lambda$,
$m_{H^\pm}=2\Lambda$.   $\sqrt{s}=320$GeV. The heavy Higgs spectrum distortion grows with $\Lambda$.} 
\epsfig{file=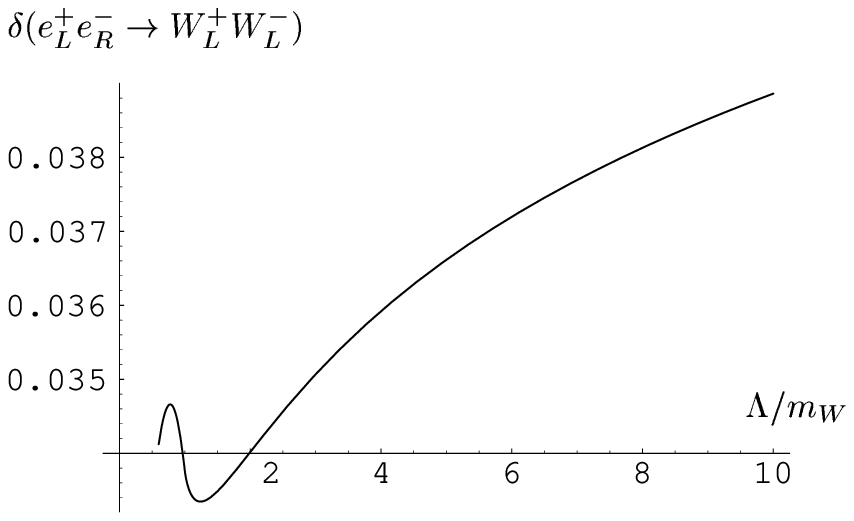, width=8cm}  
\end{figure}

The result displayed in Fig.\ref{obr1} is obtained within the following option: 
\begin{eqnarray}
\lambda_6=\lambda_7=0 & & \qquad m_{12}=0 \label{setup1}
\end{eqnarray}
Notice that in such a case the heavy Higgs limit does not exist and therefore it is 
very natural to seek for non-decoupling effects in such scenario. 
We have parametrized the magnitude of the heavy Higgs spectrum distortion by an overall
multiplicative mass scale $\Lambda$. 
Due to the relations (\ref{masses}) and (\ref{lambdas})
one obtains 
 $$
\lambda_{12}=\frac{m_{h^0}^2+m_{H^0}^2}{v^2} 
$$
In a similar manner, we can translate the $\lambda_4$, $\lambda_5$ :
\begin{eqnarray}
m^{2}_{H^{\pm}}&=&-\frac{1}{2}(\lambda_{4}+\lambda_{5}^{R})v^{2} \nonumber \\
m^{2}_{A^{\pm}}&=&-\lambda_{5}^{R} v^{2} \nonumber
\end{eqnarray}
The remaining parameters must obey
\begin{eqnarray}
\cos^2(\alpha-
\beta)& =  &\frac{1}{m_{H^0}^2-m_{h^0}^2} [(m_{H^0}^2+m_{h^0}^2)(1-2 s^2_\beta c^2_\beta)+\nonumber  \\
& & +2 s^2_
\beta c^2_
\beta(\lambda_3 v^2-2m_{H^\pm}^2)-m_{h^0}^2] \label{cosine}
\end{eqnarray}
Choosing for simplicity\footnote{Strictly speaking the limit $\beta \to \pi/2$ is unphysical making $\tan
\beta$ infinit. What we mean is rather the setup with large $\tan\beta$ motivated by low-energy SUSY
models. However, since $\lambda_6$ and $\lambda_7$ are put to zero by hand, the would-be singular 
behaviour of $\tan \beta$ is screened and does not enter the relations of our interest.} 
$$
\lambda_3=1 \qquad {\rm and}\qquad 
\beta\to\frac{\pi}{2}$$
the last unspecified parameter $\alpha$ is given by (\ref{cosine}). Let us note that there is in fact no
ambiguity
in 'fixing' $\alpha$ this way because it enters all the relevant formulae via $\cos^2(\alpha-\beta)$
only.
\\
\mbox{}\\
{$\bf e^+_Le^-_R\to W^+_L W^-_L$:}
\\
\mbox{}\\
The situation is particularly simple in the case of this polarization configuration because $\delta$ then
turns out to be independent of $t$ and $u$.
Moreover, in the case of the longitudinally polarized vector bosons in the final state we can partially
compare our results with the estimates \cite{Kanemura:1997wx} obtained by means of
the equivalence theorem.

The relevant invariants ${\cal M}_2^-$, ${\cal M}_3^-$ and ${\cal M}_5^-$ read in this case~:
\begin{eqnarray}
{\cal M}_2^- & =  & \frac{s-2m_W^2}{2 m_W^2}\sqrt{u\,t-m_W^4}
\nonumber \\
{\cal M}_3^- & =  &  \frac{s}{ m_W^2}\sqrt{u\,t-m_W^4}
\nonumber \\
{\cal M}_5^- & =  & \frac{s(s-4m_W^2)}{4 m_W^2}\sqrt{u\,t-m_W^4}
\nonumber
\end{eqnarray}
Notice that the common $\sqrt{u\,t-m_W^4}$ factors cancel in the formula
(\ref{central_relation}) and what remains is $t$ and $u$ independent. Taking into account the form of
the SM tree-level amplitude (\ref{SMtreeamplitude}), we see that the only relevant dynamical quantity in
the game is $s$ and therefore the deviation of THDM and SM predictions is 'isotropic' (at least at the
leading order).

The formulae for the formfactors $\Delta\Pi^{\gamma,Z WW}_{1,2,6}$ as well as the counterterm deviation
$\Delta\delta Z_{TGV}$ are given in \cite{Malinsky:2003bd}. Due to their enormous complexity it is
impossible to write down the results in a reasonably compact form. We have performed a numerical
simulation in {\tt Mathematica} together with the {\tt FeynCalc} and {\tt LoopTools} packages. 

The Fig.~\ref{obr1} tells us that (for a given realization of the Higgs sector) the THDM
cross section of the considered process should be enhanced by several percents with respect to its SM
value and grow
logarithmically with the heavy Higgs spectrum distortion $\Lambda$, in perfect agreement with what was
anticipated on theoretical grounds in Section~\ref{nondecoupling} and \cite{Malinsky:2003bd}. 
We have
checked that qualitatively the same happens also in other similar setups.

\subsection{\label{sect5-2}Decoupling regime} 
\begin{figure}[h]
\caption{\label{obr2}$\delta$ as a function of $\Lambda$ for $m_H^0=\Lambda+10m_W$,
$m_{A^0}=\Lambda-6m_W$, $m_{H^\pm}=\Lambda-5 m_W$.   $\sqrt{s}=200$GeV. The heavy Higgs bosons decouple
with the rising overall scale $\Lambda$ since the distortion $\Delta\sim m_{W}$ of their spectrum
remains fixed and $\Delta/\Lambda\to 0$.}
\epsfig{file=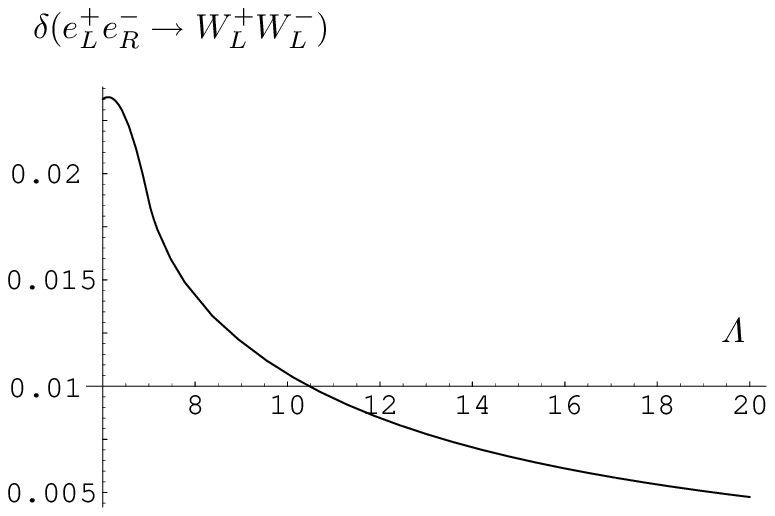,width=8cm}  
\end{figure}

We achieve the {\it decoupling regime} by taking all the heavy Higgs masses
quasidegenerate with a constant distortion $\Delta \sim m_{W}$ much smaller than the overall scale
$\Lambda$
driving the heavy part of the THDM Higgs spectrum. For $\Delta/\Lambda \to 0$ the $\delta$ should tend to
$0$.
As can be seen in Fig. \ref{obr2} it is indeed the case. This provides a nontrivial consistency check of
our results. What is interesting is the fact that this picture was obtained within the setup with
$m_{12}\to 0$.  There is nothing that would contradict our previous considerations, because $m_{12}$
{\it is not the only mass singlet parameter in the game}. There can be still big 'hidden' singlets
responsible for such behaviour, namely the $m_{11}$ and $m_{22}$ just 'translated' throught the
minimality conditions into combinations of the other parameters. This can work whenever
at least one of the $\lambda_{6}$ or $\lambda_{7}$ is nonzero (justifying the choice (\ref{setup1}) in
the  non-decoupling case). 
As we have checked, it works even better in the case of the 'apparent' decoupling setup
driven by the allowed nonzero $m_{12}$.

\section{Conclusions}
We have seen that within the THDM framework a non-decoupling behaviour of heavy Higgs bosons can occur
quite naturally. There is a simple phenomenological criterion for recognizing the
character of the heavy Higgs effects in the physical amplitudes, namely the magnitude of the distortion of
the heavy part of the Higgs spectrum. Since there is no possibility to get large deviations from the
quasidegenerate heavy Higgs spectrum in the MSSM, the heavy Higgs bosons in the minimal supersymmetry
should always decouple, which is fully compatible with the explicit analyses in the literature
\cite{Dobado:2000pw}. 

We have given a general one-loop estimate of such effects in the case of the physical process      
$e^+e^-\to W^+W^-$ within THDM. In many cases the deviation of its cross section from
the SM
results can be of the order of several percent, in good agreement with some previous partial analyses with
longitudinally polarized vector bosons in the final state based on ET approximation.
In principle, such effects could be visible in future experiments making the two Higgs doublet extension
still a viable candidate of a theory beyond the Standard Model. 
 
\section*{Acknowledgements} 
This work was supported by "Centre for Particle Physics", project No.
LN00A006 of the Ministry of Education of the Czech Republic.

\end{document}